\documentclass[fleqn,usenatbib]{mnras}

\usepackage{newtxtext,newtxmath}

\usepackage[T1]{fontenc}

\DeclareRobustCommand{\VAN}[3]{#2}
\let\VANthebibliography\thebibliography
\def\thebibliography{\DeclareRobustCommand{\VAN}[3]{##3}\VANthebibliography}

\usepackage{amsmath}
\usepackage{caption}
\usepackage{wrapfig}
\usepackage{subcaption}

\usepackage{graphicx}	
\usepackage{amsmath}	

\newcommand{\EQRef}[1]{Eq.~\ref{#1}}

\newcommand{\FIGRef}[1]{Fig.~\ref{#1}}

\newcommand{\TABRef}[1]{Table~\ref{#1}}

\newcommand{\SECRef}[1]{Section~\ref{#1}}

\title[Short title, max. 45 characters]{GPU accelerated MHD in the DISPATCH framework using directive-based programming}

\author[M. Haahr et al.]{
Michael Haahr,$^{1,2}$\thanks{E-mail: michhaa@uio.no}
Troels Haugb{\o}lle,$^{2}$
{\AA}ke Nordlund$^{1,2}$,
Sven Karlsson$^{3}$,
and Eloi Martaillé Richard$^{1}$
\\
$^{1}$Institute of Theoretical Astrophysics, University of Oslo, Sem Sælands Vei 13, 0371, Oslo, Norway\\
$^{2}$Niels Bohr Institute, University of Copenhagen, Øster Voldgade 5, DK-1350 Copenhagen, Denmark\\
$^{3}$Technical University of Denmark (DTU) Richard Petersens Plads, 322, 106, 2800 Kgs. Lyngby, Denmark
}

\pubyear{\the\year{}}

\begin{document}
\label{firstpage}
\pagerange{\pageref{firstpage}--\pageref{lastpage}}
\maketitle

\begin{abstract}
We present a GPU-accelerated implementation of a magnetohydrodynamic (MHD) solver using directive-based programming with OpenMP target offloading. The solver is integrated into the DISPATCH framework, which organises the computational domain into a collection of asynchronously updated patches. To reduce GPU kernel launch overhead, patches are grouped into 'bunches' that are updated collectively. While porting the particular solver required a complete code refactoring, it yielded performance gains on both GPU and CPU. A stand-alone mini-app achieved a 7.3× speed-up compared to a single NVIDIA A100 GPU to seven AMD 7F72 Rome CPU cores. Within the full DISPATCH framework, the GPU-accelerated MHD Bunch solver showed excellent agreement with the CPU-based reference implementation on standard test problems such as the Sod shock tube and Orszag–Tang vortex. In large-scale 3D tests, the GPU implementation achieved a 9.8× overall speedup, comparing one GPU to 12 CPU cores, with the core MHD update routine being two orders of magnitude faster on the GPU than on a single CPU core. These results demonstrate that OpenMP offloading can provide substantial performance improvements for astrophysical codes while maintaining portability and accuracy. The work also demonstrates how new codes should be structured to allow simple and efficient directive-based GPU offloading.
\end{abstract}




\section{Introduction}

In the past decade, a growing fraction of the computational capacity at large supercomputers has been based on accelerators, such as GPUs. Nonetheless, transitioning simulation frameworks to execute on these accelerators efficiently has proven more difficult, and the majority of astrophysical modelling frameworks cannot utilise GPUs. Historically, numerical codes used in astrophysical simulations have pushed the limits of available computer hardware, as many astrophysical phenomena cannot be reproduced in a controlled laboratory environment. Therefore, \textit{in-silico} modelling has been instrumental in understanding which physical forces are at play, developing new theory, and interpreting observations. The creation of digital twins has gained traction in interpreting observations better. Since NVIDIA released CUDA in 2007 \citep{GPU_definition}, using GPUs for general-purpose computing has become increasingly popular, but it is not yet common practice for astrophysical simulations. In modern supercomputers, such as LUMI\footnote{\url{https://www.lumi-supercomputer.eu/}}, over 90\% of the theoretical floating point operations per second (FLOPS) are provided by GPUs. To continue pushing the limits of computer hardware, astrophysical simulations on such exascale systems have to utilise the compute resources available in GPUs.

GPU hardware differs from CPU hardware and requires a different programming approach.
Modern CPUs are designed for fast serial execution and have up to a few hundred cores, each with supporting logic for features such as out-of-order execution, branch prediction, and memory prefetch. GPUs can instead have more than 10,000 cores, and a significant fraction of the silicon is dedicated to execution units and registers to hold the data. The cores are scheduled according to the \textit{single instruction multiple threads} (SIMT) execution mode, with many cores executing the same instruction on multiple data, and multiple thread groups executing on a single set of cores, to mask latency in fetching the large amount of data required to keep the execution units busy. The cores in the collection all execute the same code, with the only difference being their unique thread ID. This thread ID is commonly used in place of an index in a traditional for loop, which alters the code's structure. In addition to this, the GPU and CPU have separate main memory\footnote{This is currently changing with the arrival of cache-coherent or single memory hierarchies in, e.g., NVIDIA Grace-Hopper and AMD MI300A GPUs.}, and the data transfer between GPU and CPU must be explicitly handled by the programmer, or implicitly by the programming framework. Due to the complexity of writing explicit GPU code, the use of compiler directives has become increasingly popular.

Directives are comments in the source code that the compiler can read and use to compile the code in a different manner. 
An 'offload directive' may be placed at the beginning of a for loop to instruct the compiler to compile the loop as GPU code. The compiler will rewrite the loop, handle the data transfer between the GPU and the CPU, and distribute the loop iterations as work items for different threads on the GPU. Since the directives are only comments, the main code can remain unchanged, which speeds up the porting process and helps maintain a single code base that supports different hardware architectures. The primary drawback of the compiler-based approach is the limited compiler support, which can render porting impossible for certain code.

Accelerating codes with GPUs is an active area of research. The GenASIS code \citep{GPU_SUMMIT} was accelerated using a combination of Fortran and C. FARGO3d \citep{fargo3d} is a fully GPU-based framework written solely in CUDA, with partial support from a Python-based auto-translator tool that generates CPU or GPU code based on a template. The \texttt{Kokkos} library, which provides high-level abstractions for parallel execution and memory management, has also gained popularity over the last decade as a means to accelerate numerical codes while maintaining performance portability across diverse architectures. Recently, several approaches for GPU acceleration were investigated in the GENE code \citep{Fusion_GPU}, where the authors ultimately chose to use CUDA C++. The GENE \citep{Fusion_GPU} effort explored OpenACC and OpenMP but did not pursue them due to limited compiler support at the beginning of their project. Similar conclusions were drawn in the early stages of this project; however, with GCC 10.2, OpenMP offload in the GCC compiler matured sufficiently to support the code. OpenMP offload features have previously been successfully tested; however, the results obtained have all been on small mini-apps \citep[e.g.][]{OpenMP_Proxy}. To our knowledge, this is the first time a full-scale astrophysics code has been ported to GPU using only OpenMP directives.

If a directive-based porting is generally viable, it could enable more codes to utilise the GPU. Even though a directive approach, in principle, does not require rewriting the code, as we show below, it may be necessary in practice to achieve better performance and for the code structure to map well to the limitations imposed by the hardware. With compiler directives, one can, in principle, always offload computations to GPUs; however, based on compiler support and user implementation, the resulting code may perform worse than the original. For the porting to be a success, it must also provide a satisfactory performance increase. Below, we investigate if, and how, a directive-based approach can be used to successfully port the MHD solver in the DISPATCH framework to a GPU.

The remainder of this article is structured as follows. In \SECRef{sec:methods}, we describe the numerical methods and the DISPATCH framework, followed by the porting strategy and integration of the GPU-accelerated solver. \SECRef{sec:results} presents validation tests and performance benchmarks comparing the GPU and CPU implementations. In \SECRef{sec:discussion}, we discuss the implications of the results, identify current limitations, and explore avenues for future improvements. Finally, \SECRef{sec:conclusion} summarises our findings and outlines prospects for extending GPU acceleration within the DISPATCH framework.

\section{Methods}
\label{sec:methods}
\subsection{Numerical Methods}

As a background for considerations on porting MHD solvers to GPU, we present the numerical methods used for solving the MHD equations in this section. The equations of interest in this paper are the conservative formulations of ideal MHD:
\begin{align}
    \frac{\partial \rho}{\partial t} + \nabla \left[ \rho \mathbf{u} \right] & = 0 \\
    \frac{\partial \rho \mathbf{u}}{\partial t} + \nabla \left[ \mathbf{u} \otimes \mathbf{u} + P \mathcal I - \mathbf{B} \otimes \mathbf{B} \right] & = 0 \\
    \frac{\partial E}{\partial t} + \nabla \left[ (E + P) \mathbf{u} - (\mathbf{B} \cdot \mathbf{u}) \mathbf{B} \right] & = 0 \\ 
    \frac{\partial \mathbf{B} }{\partial t} + \nabla \times  \left( \mathbf{u} \times \mathbf{B} \right) & = 0\,,
\end{align}
where bold characters indicate vectors, $\mathbf{B}$ is the magnetic field, $\mathbf{u}$ is the velocity vector, $\rho$ is the mass density, $E$ is the total energy density, $P = P_\textrm{gas} + P_\textrm{mag}$ is the sum of the gas and magnetic pressure, and $\mathcal I$ is the identity matrix. This set of equations must be complemented by an equation of state that links the density, pressure, internal energy and temperature.
The equations may be written in conservative form as a single vector equation
\begin{equation}
\mathbf{U} + \nabla \cdot \mathbf{F(U)} = 0\,,
\end{equation}
where $\mathbf{U}$ is a so-called \emph{state vector} of conserved variables and $\mathbf{F}$ their corresponding fluxes. In 1D along the $x$-axis it becomes
\begin{equation}
\label{eq:MHD_vector}
\mathbf{U}=
\begin{bmatrix}
\rho \\
\rho u\\
\rho v\\
\rho w\\
E \\
B_x\\
B_y\\
B_z
\end{bmatrix}, \,
\mathbf{F(U)}=
\begin{bmatrix}
\rho u \\
\rho u^2 + P - B_x^2\\
\rho v\,u - B_y\,B_x\\
\rho w\,u - B_z\,B_x\\
u(E+P) - B_x(\mathbf{u} \cdot \mathbf{B}) \\
0\\
B_y\,u-B_x\,v\\
B_z\,u-B_x\,w
\end{bmatrix}\,,
\end{equation}
with the velocity and magnetic field written out in components $\mathbf{u}=(u,v,w)$ and $\mathbf{B}=(B_x$,$B_y$,$B_z)$. Assuming an adiabatic equation of state with adiabatic index $\gamma$, the total energy is related to the gas pressure as $P_\textrm{gas} = (\gamma-1)(E-\frac{1}{2}\rho\mathbf{u}^2 - \frac{1}{2}\mathbf{B}^2)$. To solve this set of PDEs numerically, we use a flux-based finite-volume approach that explicitly conserves mass, momentum and total energy. To maintain the solenoidal constraint $\nabla \cdot \mathbf{B} = 0$, the electric and magnetic fields are 'staggered' as illustrated in \FIGRef{fig:B_and_EMF}, and the magnetic fields are updated using constrained transport. At the interface between each cell, a discontinuity will occur. Such a discontinuity gives rise to a Riemann problem \citep{riemann_book}. The numerical methods for solving Riemann problems have evolved over the years, and in DISPATCH, several approximate Riemann solvers are maintained.

\subsubsection{HLL Solver}
Harten, Lax and van Leer proposed an approximate solver to the Riemann problem---the HLL solver \citep{HLL_original}. The idea behind this general type of solver is to separate the interface into different regions. The regions are separated by eigenmodes propagating with characteristic speeds. In the HLL solver, three regions exist. The initial Left and Right regions are described by the state vectors $\mathbf{U_L}$ and $\mathbf{U_R}$, and the central star-region with state vector $\mathbf{U^*}$. This corresponds to the light grey shaded area in \FIGRef{fig:HLLD}. The state vector within the star region is constructed as an integral average of the fluid state inside this region and can be calculated given the left and right states. Depending on the velocities of the separating eigenmodes, either of the three regions may occupy the interface $x=0$ in the figure. The values given by this region are used to calculate the flux going through the interface using equation \ref{eq:MHD_vector}. The HLL solver is very stable, but because the state vector and related flux in the star region are based on the integral average of the solution inside the region, the solution is quite diffusive.

\subsubsection{HLLD Solver}
\begin{figure*}
\centering
  \includegraphics[width=0.95\textwidth]{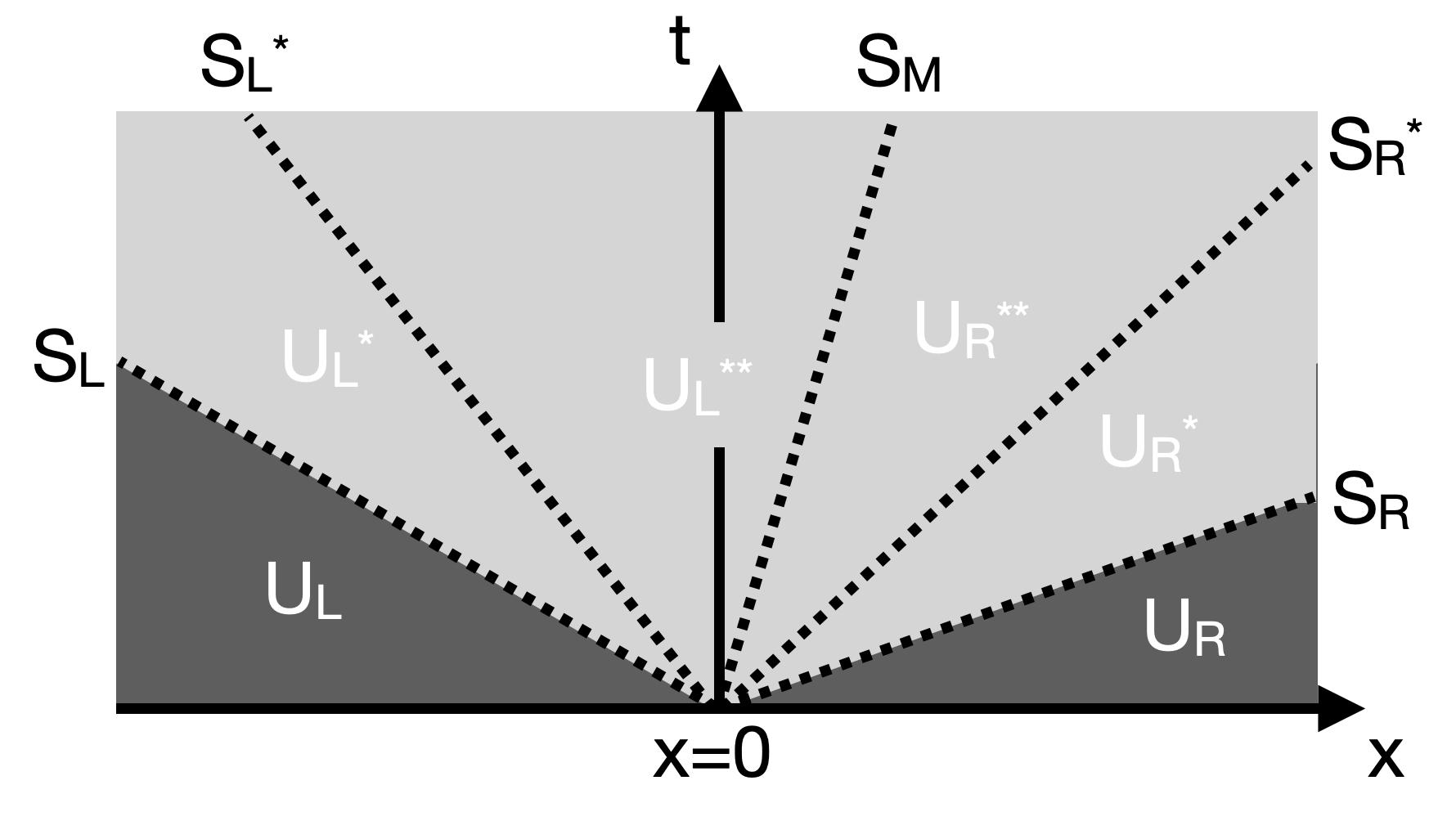}
\caption{Six-state HLLD solver for MHD. The initial values are given to the left ($U_L$) and right ($U_R$). The four intermediate states and three intermediate waves are calculated based on the right/left data and the estimated wave speeds $S_L$ and $S_R$. The light grey shaded area corresponds to the single intermediate state in the HLL solver.}
\label{fig:HLLD}       
\end{figure*}

The HLLD solver is an extension of the HLL solver that better models discontinuities and magnetic fields. The HLLD solver separates the fluid into six regions, as shown in figure \ref{fig:HLLD}. For a full explanation of these regions, we refer to \citet{HLLD_2005}. The basic idea is the same as in the HLL solver, but with more eigenmodes motivated not only by conservation properties, but also by the specific structure of the MHD system. In figure \ref{fig:HLLD}, the region at the interface is the left double star region ($\mathbf{U_L^{**}}$). Using this solver works well for the hydrodynamic variables, but a simple flux-based approach does not ensure that the magnetic field remains divergence-free.

\subsubsection{Constrained transport and 2D-extension of the HLLD solver}
\begin{figure*}
    \centering
    \includegraphics[width=0.9\textwidth]{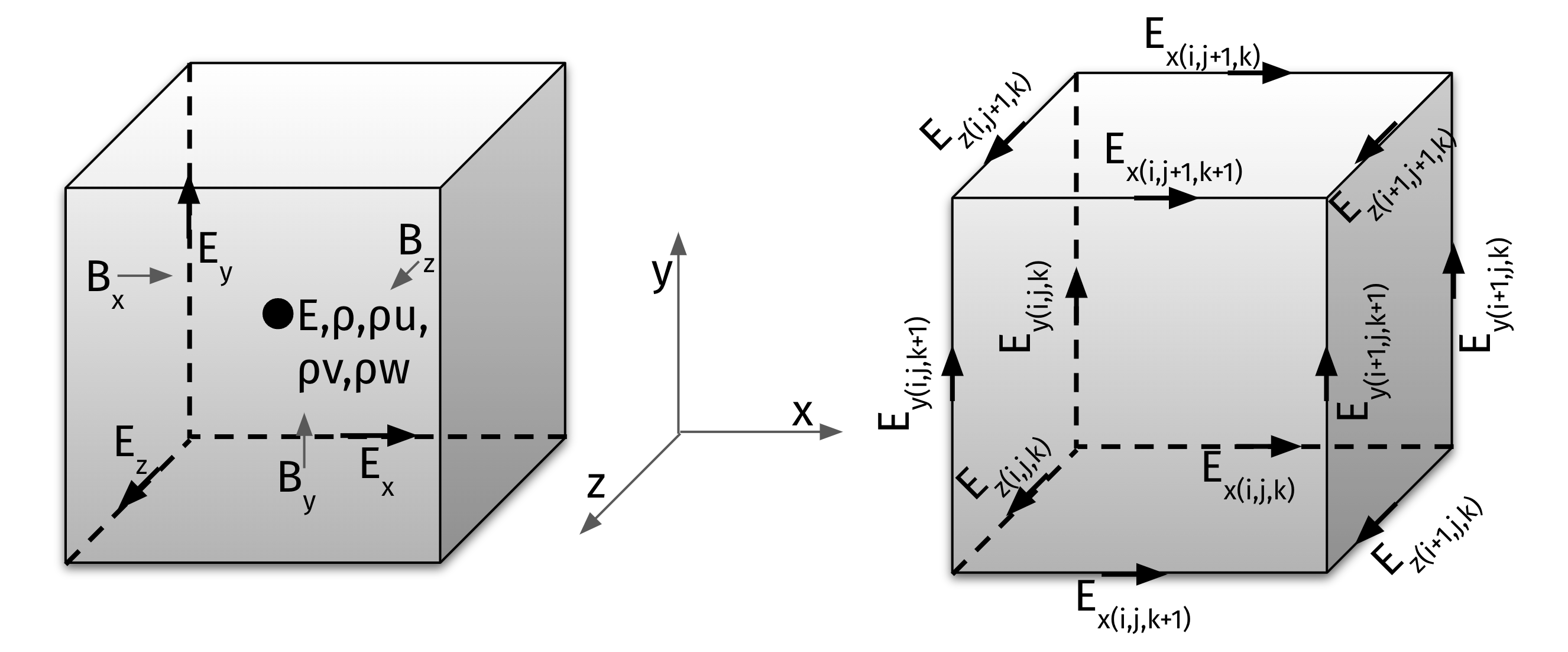}
    \caption{Left Panel shows the location of the primitive variables (centre), magnetic field (face), and edge-averaged EMF (edge).  
    The right panel shows the edge-averaged EMF on all edges of the cell. Most of the EMFs are stored in neighbouring cells, as indicated by the array index subscripts.}
    \label{fig:B_and_EMF}
\end{figure*}

Previously, the magnetic field was typically kept divergence-free by adding a divergence-cleaner step or by reformulating the MHD equations \citep[see][]{Divergence_cleaning_1998,BRACKBILL1980426}. \citet{HLLD_2012} instead proposed extending the HLLD to include the induction equation in a constrained transport formulation. The magnetic fields at each cell interface can be updated using the edge-centred averaged EMF. With the assumption of infinite conductivity in ideal MHD, Ohm's law simplifies to $\mathbf{E} = - \mathbf{u} \times \mathbf{B}$. By calculating the EMF and then using the reduced Ohm's law to update the magnetic field, one ensures that the magnetic field remains divergence-free. The EMF is calculated so that it is located at the edges of the cell as illustrated in \FIGRef{fig:B_and_EMF}. The EMF in the $z$ direction, for example, is calculated by the following equations: 
\begin{equation}
    {E_z^n}_{(i,j,k)} = \Bar{v_x}\Bar{B_y} - \Bar{v_y} \Bar{B_x}
\end{equation}
with 
\begin{equation}
    \Bar{v_x} = \frac{1}{4}({v_x^n}_{(i,j,k)}+{v_x^n}_{(i,j-1,k)}+{v_x^n}_{(i-1,j,k)}+{v_x^n}_{(i-1,j-1,k)})
\end{equation}
\begin{equation}
    \Bar{v_y} =\frac{1}{4}({v_y^n}_{(i,j,k)}+{v_y^n}_{(i,j-1,k)}+{v_y^n}_{(i-1,j,k)}+{v_y^n}_{(i-1,j-1,k)})
\end{equation}
\begin{equation}
    \Bar{B_x} = \frac{1}{2} ({B_x^n}_{(i,j,k)}+{B_x^n}_{i,j-1,k})
\end{equation}
\begin{equation}
    \Bar{B_y} =\frac{1}{2} ({B_y^n}_{(i,j,k)}+{B_y^n}_{i-1,j,k})
\end{equation}
where the subscripts $i$, $j$, and $k$ indicate the array index position in the $x$, $y$, and $z$ dimensions of each variable, and the superscript indicates time. 
The EMF is calculated for time $t+\frac{1}{2}\Delta t$, and is used to calculate the update of the magnetic field. For $B_x$ this would be:
\begin{align}
\label{eq:EMF_to_B}
{B^{n+1}_x}_{(i,j,k)} =\; & {B^{n}_x}_{(i,j,k)} \nonumber \\
&+ \frac{{E^{n+\frac{1}{2}}_z}_{(i,j+1,k)} - {E^{n+\frac{1}{2}}_z}_{(i,j,k)}}{dy} \, dt \nonumber \\
&- \frac{{E^{n+\frac{1}{2}}_y}_{(i,j,k+1)} - {E^{n+\frac{1}{2}}_y}_{(i,j,k)}}{dz} \, dt
\end{align}

\begin{figure*}
\centering
  \includegraphics[width=0.85\textwidth]{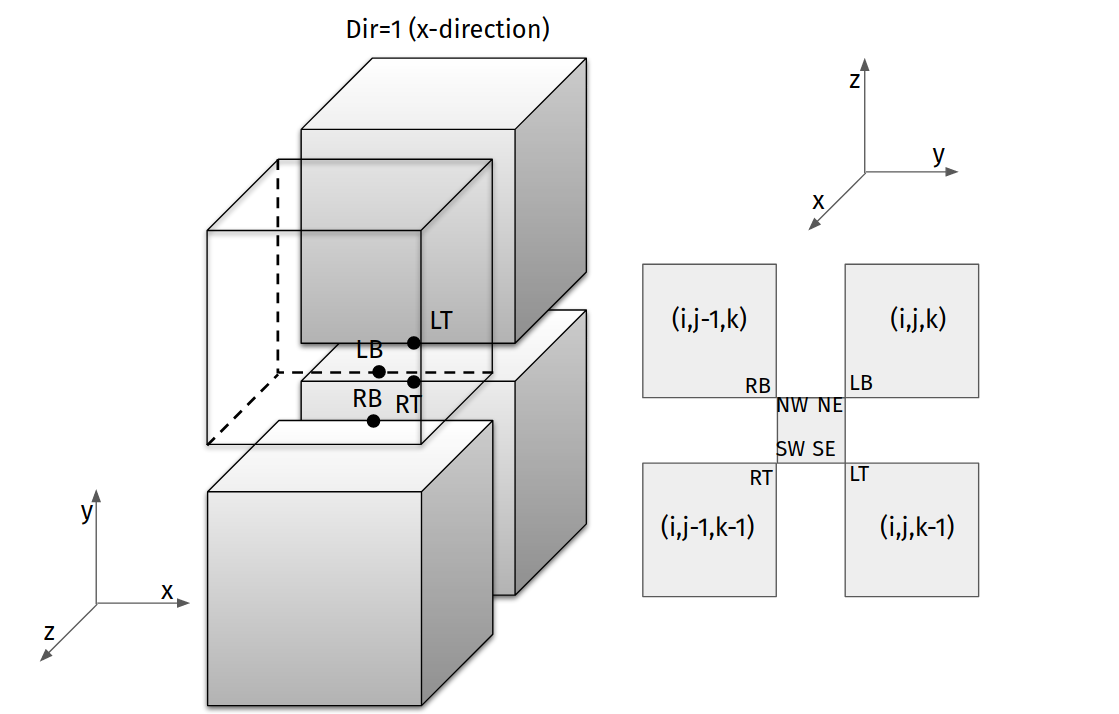}
\caption{Cell corner values to edge-centred corner values in the $x$-direction. SW, SE, NW, and NE values are used to solve the 2D Riemann problem and calculate the EMF. LB, LT, RB, and RT values are the interpolated corner values in each cell.}
\label{fig:corner_values}
\end{figure*}

As discussed in the next section, in the numerical scheme, values are first predicted at time $t+\frac{1}{2}\Delta t$, and these predicted values are then used in the Riemann problem.
Predicting the edge-centred values in four neighbouring cells gives rise to a 2D Riemann problem that must be solved to estimate the EMF properly. This is illustrated in \FIGRef{fig:corner_values}. Values are interpolated for the LeftBottom (LB), RightBottom (RB), LeftTop (LT), and RightTop (RT) corners in each direction. These values are renamed in the 2D Riemann solver (HLLD 2D hereafter) as SW, NW, SE, and NE values to highlight that these are structured differently. 
The HLLD 2D method ensures that the magnetic field remains divergence-free.

\subsubsection{The MUSCL Scheme}
The HLLD 2D Riemann solver is conservative, but numerical fluctuations may still make the solver unstable.
It can be shown \citep[see][for details]{riemann_book} that the coupled PDEs in \EQRef{eq:MHD_vector} can be expressed instead as a set of decoupled PDEs.
The new set of PDEs operates on a different set of variables: density, velocities, pressure, and magnetic fields. In the MUSCL scheme, these are referred to as the primitive variables. 

The MUSCL scheme is used for extrapolating interface values needed by the Riemann solvers and updating the cell-centred values in a leapfrog-like method. First, the primitive variables are calculated, and interface values are constructed based on slope (gradient) estimates in each cell. The choice of slope estimate is a crucial step to prevent or suppress artificial oscillations in the solutions.

Once the gradients are found, predictions are made for the time $t+\frac{1}{2}\Delta t$. This is done purely using existing values and the gradients, and is based on a simple forward-in-time Euler discretisation of the primitive variables. The predictions for the magnetic field are calculated based on the EMF as previously discussed. The predictions are then used in the HLLD and HLLD 2D solvers to compute the interface fluxes that are used to advance the conservative variables in time. 

Given a physically correct initial condition, the MUSCL scheme ensures that the solution remains physical. Optimising the MUSCL scheme is key for running MHD code efficiently, but it is not enough to ensure good performance on modern supercomputers. To ensure good performance, we need a scalable framework.

\subsection{DISPATCH}
There currently exist many different codes for astrophysical fluid simulations. Some examples are ZEUS \citep{ZEUS}, FARGO3D \citep{fargo3d}, BIFROST \citep{bifrost}, STAGGER \citep{stagger}, and GenASIS \citep{GenAsis}. These codes are, at their core, grid-based. Traditional codes like these only scale well up to a limit on the number of cores used. One might partially compensate for this by custom-fitting a code specifically to the system on which it is to be run. However, as discussed in \citet{framework_limit}, modern codes are too large to be realistically modified to run optimally on a given machine or architecture.

DISPATCH \citep{Dispatch_framework} is a framework written in Modern Fortran that offers practically unlimited scalability. DISPATCH breaks with the traditional domain-decomposition strategy. Instead of statically splitting up a grid into relatively large chunks, smaller chunks are used and are represented as Fortran objects, referred to as \emph{extended data types}. These hold several \emph{time slices} of the grid values, as well as metadata such as location, mesh information, time, and timestep. These chunks are referred to as \emph{patches} in DISPATCH, and are one type of extension to a basic, underlying task concept. 

DISPATCH is written as a combination of different modules that are, in principle, independent of each other. The handling of boundary conditions, task scheduling, load-balancing, etc.\ is thus not dependent on the type of solver. Therefore, when implementing new solvers in DISPATCH, one only has to ensure that the solver works for a single patch. The framework will ensure that it works effectively when running more complex setups with thousands of patches across multiple compute nodes. This enables solvers to be implemented and tested outside the framework.

\subsection{HLLD Mockup}
To test porting strategies, a mockup solver was created. First, the simple HLL scheme was implemented in a mockup and various porting strategies were tested. We evaluated both OpenACC and OpenMP versions. Since the development of OpenACC has been taken over by NVIDIA, concerns arose regarding the future portability of an OpenACC implementation. At the same time, LUMI announced that it would deploy AMD GPUs. We therefore chose to use OpenMP as our porting scheme.

We found that the most effective strategy was to ‘bunch’ many patches together and update each step in the MUSCL routine across all patches.

The full HLLD 2D solver was then implemented. The HLLD solver is an almost verbatim copy of the RAMSES HLLD solver \citep{Ramses}. In DISPATCH, this solver is referred to as the RAMSES/HLLD solver. The bunching strategy was also adopted here, but some issues arose. The entire MUSCL scheme appeared too complex for the compilers to offload, resulting in GPU memory errors. It was ultimately necessary to split each part of the MUSCL routine into individual kernels. Specifically, the monolithic \texttt{trace3D} routine, which is central to the RAMSES code, needed to be divided into multiple kernels.

Memory errors also appeared when temporary arrays were allocated and deallocated during the transfer of a bunch to the GPU. It seems that the GCC compiler does not yet reliably deallocate arrays on the GPU, resulting in a gradual increase in memory usage. We therefore opted to use statically allocated work arrays. This resolved the error and resulted in a slight improvement in overall performance. Although the arrays are now statically stored, the memory footprint remains effectively unchanged in realistic scenarios where all bunches are either in use, being prepared, or finalising execution.

Two potential bugs were identified in the RAMSES solver. First, the RB and LB values were inadvertently flipped in the $y$-direction when calculating corner values. However, since this was done twice in reverse order, the errors effectively cancelled each other out. While this had no effect on the final results, it did compromise intermediate quantities. Second, for the EMF computed in HLLD 2D, the values were not multiplied by the correct $dx$, $dy$, or $dz$. In DISPATCH/RAMSES, $EMF_x$ is multiplied by $dt/dy$, $EMF_y$ by $dt/dz$, and $EMF_z$ by $dt/dx$. This is done under the assumption that cell sizes are uniform in all directions. However, if the cell sizes are not identical, numerical discrepancies will arise. In practice, cell sizes in DISPATCH are typically chosen to be cubic. Still, errors could emerge in 2D test cases if the collapsed dimension had a nominal cell size differing from the active directions.

In the end, the kernels were gradually merged, and by restructuring the array layout of the variables and setting the OpenMP scheduling scheme to \texttt{schedule(static)}, all operations could be encapsulated within a single offload region, giving the best performance. Here, a “kernel” refers to any code block enclosed between \verb|!$omp target| and \verb|!$omp end target| directives. Within this region, multiple \verb|!$omp parallel do| blocks are employed to parallelise different parts of the update. Once this configuration was functioning correctly, the solver was integrated into the DISPATCH framework.

\subsection{DISPATCH Integration}
The solver itself was integrated with minimal effort, although a new bunching module had to be added to support the bunching scheme. This module, \texttt{offload\_mod}, operates by placing patch data ready for update into a temporary work array. Such an array of tasks is referred to as a \emph{bunch}. Once a bunch reaches a certain size, all patches within it are offloaded for GPU execution. The new solver that supports this bunching module is referred to as \texttt{MHD\_Bunch}.

\begin{figure*}
\centering
  \includegraphics[width=0.9\textwidth]{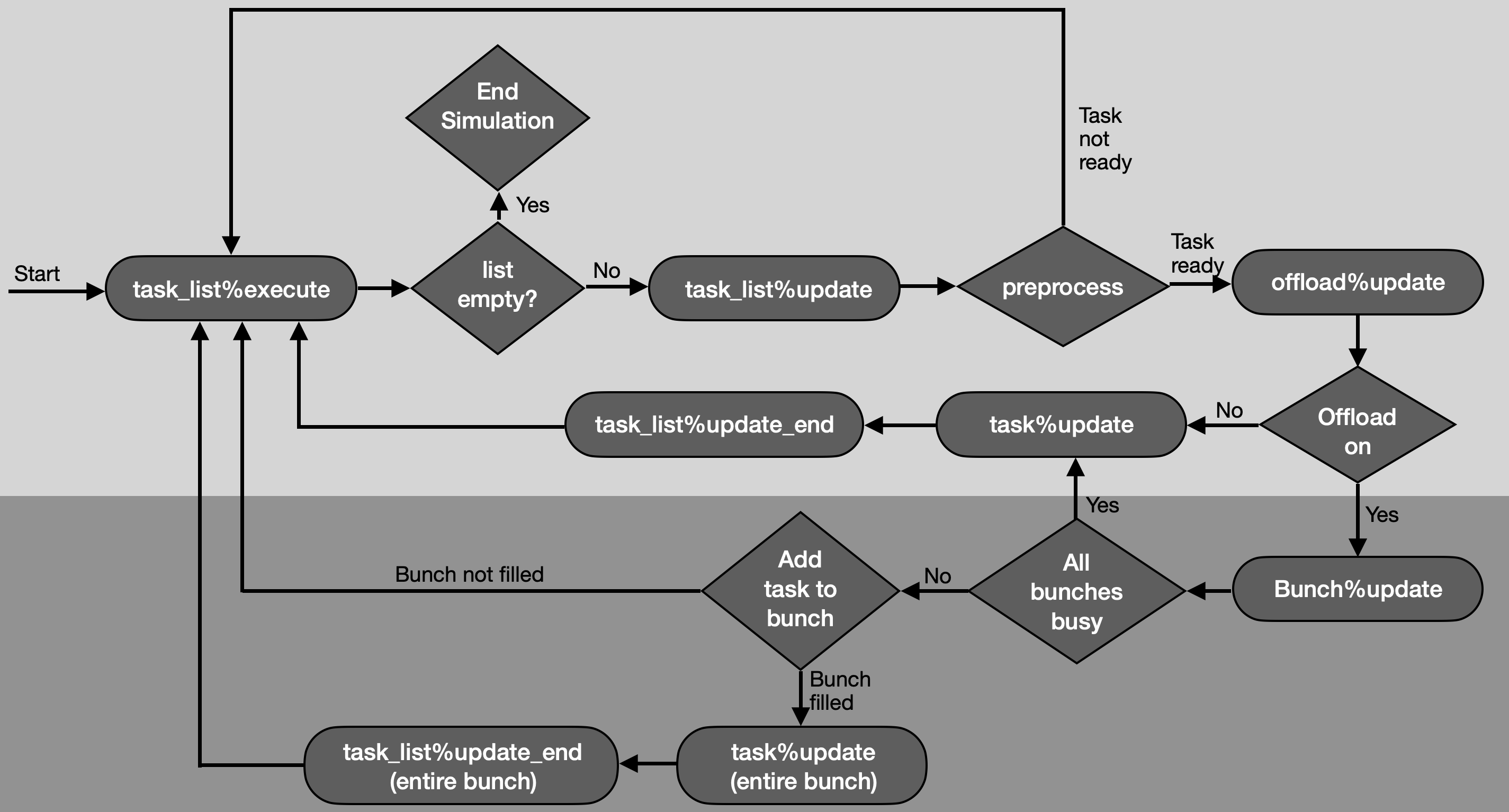}
\caption{Flowchart showing the basic execution flow of the DISPATCH framework with the bunching module. The light-shaded area shows the previous execution flow, and the dark-shaded area shows the added steps to implement bunching.}
\label{fig:flowchart_bunch}
\end{figure*}

The offload module can contain multiple bunches, which may be filled in a round-robin fashion to conceal pre- and post-processing work as well as memory transfers. The solver can also be run in CPU-only mode, or with most updates offloaded to GPUs while some are handled by CPUs. Although not yet implemented, it is in principle possible to distribute the bunches across multiple GPUs on a single MPI rank. However, we believe assigning a separate MPI rank to each GPU may be a more effective approach. A simplified flowchart illustrating the new and original DISPATCH execution flow is shown in \FIGRef{fig:flowchart_bunch}.

The performance and correctness of the \texttt{MHD\_Bunch} solver were tested and compared to those of the existing solver, with the results presented in the sections below.

\subsection{HLLS Solver}
While not the primary focus of this article, the bunching strategy has also been applied to the recently developed HLLS solver, which is under active development. As the development of this solver is still ongoing, we will not discuss it in detail. Briefly, the HLLS solver is a modified version of the HLLD solver developed by \citet{hlls_solver}. It follows the same principles as the HLLD solver but replaces total energy with entropy as the ‘conserved’ variable. The motivation for this substitution is that in some contexts—such as the convection zones of stars—it is important to transport entropy per unit volume in a conserved manner, while dissipative terms can be treated as explicit source terms.

\subsection{Mockup}
\begin{figure*}
    \centering
    \includegraphics[width=0.9\textwidth]{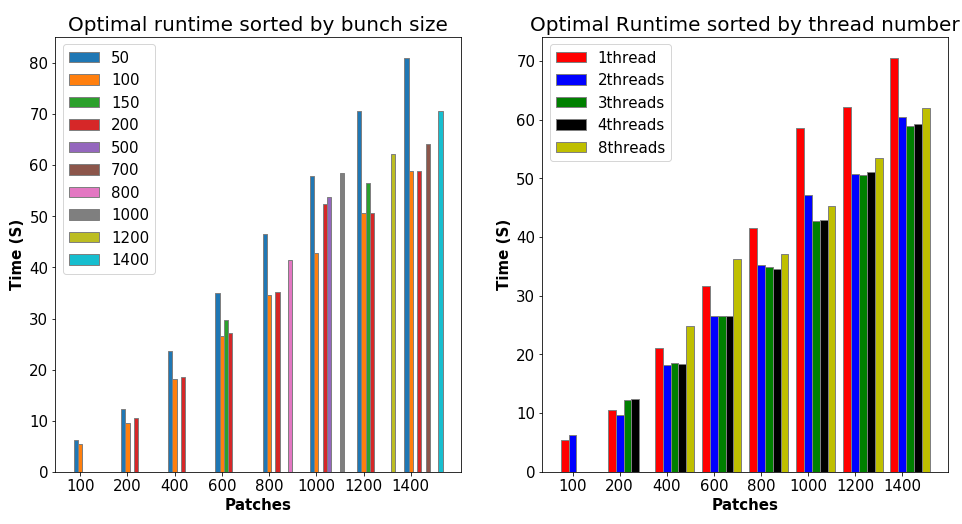}
    \caption{Optimal run-times, sorted by bunch size and number of threads for different total patch counts. Lower values indicate better performance.}
    \label{fig:bunch_thread_times}
\end{figure*}

The speed-up was first examined using the HLLD mockup. We initially investigated the impact of bunch size; the results are shown in \FIGRef{fig:bunch_thread_times}. The simple mockup bunching approach had a limitation in that only bunch sizes divisible by the total number of patches could be chosen, so not all combinations were testable. However, the results consistently show that a bunch size between 100–400 and a thread count between 3–4 yields optimal performance. These experiments were conducted on an A100 GPU, which has 108 streaming multiprocessors (SMs), which are GPU cores grouped for instruction-level parallelism. We therefore conclude that the optimal bunch size is around 1–4 times the number of SMs in the target GPU. Furthermore, the results clearly show that using more than one thread helps to hide kernel launch overhead.

\begin{figure*}
    \centering
    \includegraphics[width=\textwidth]{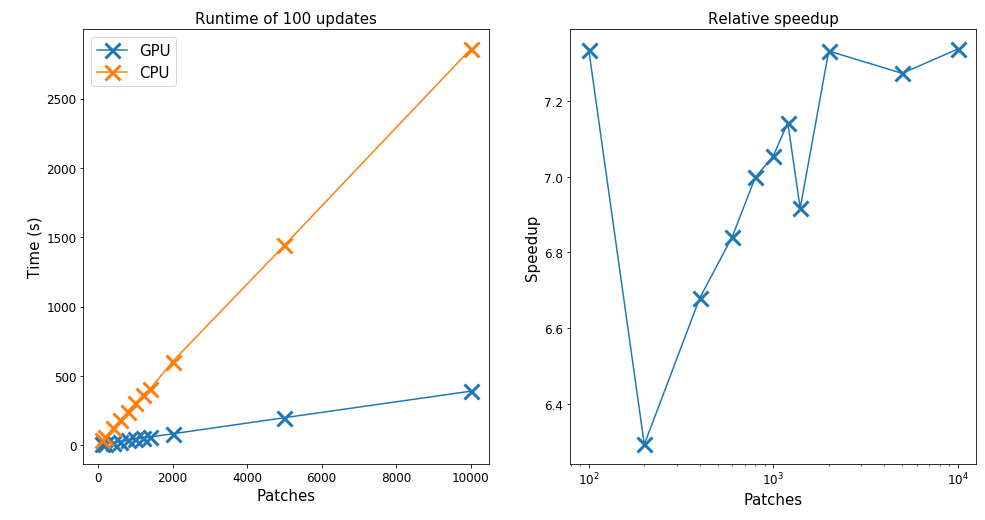}
    \caption{Runtime and speed-up comparing one GPU to seven CPU threads.}
    \label{fig:7threads}
\end{figure*}

The runtime of the CPU and GPU versions of the mockup was then tested for a large number of patches. The results are presented in \FIGRef{fig:7threads}. We chose to compare the performance of the mockup on one GPU with seven CPU cores, as this configuration approximates the setup used on Summit at the Oak Ridge Leadership Computing Facility\footnote{\url{https://www.olcf.ornl.gov/summit/}}, allowing for easy comparison with other codes tested on that system. Except when running with just 100 patches, the speed-up increases from approximately 6.4 to 7.3 and then levels off. We believe the high speed-up observed when running 100 patches arises from the A100's 108 SMs, which allow the bunch to nearly saturate the GPU. Similar experiments were run using 5, 10, 20, and 40 threads. In these cases, the speed-up for 10,000 patches was found to be 9.3, 4.9, 3.0, and 3.0, respectively. With the mockup showing promising performance improvements, the solution was integrated into DISPATCH and validated.

\section{Validation and Performance}\label{sec:results}

\subsection{DISPATCH Validation}
\begin{figure*}
\centering
  \includegraphics[width=0.9\textwidth]{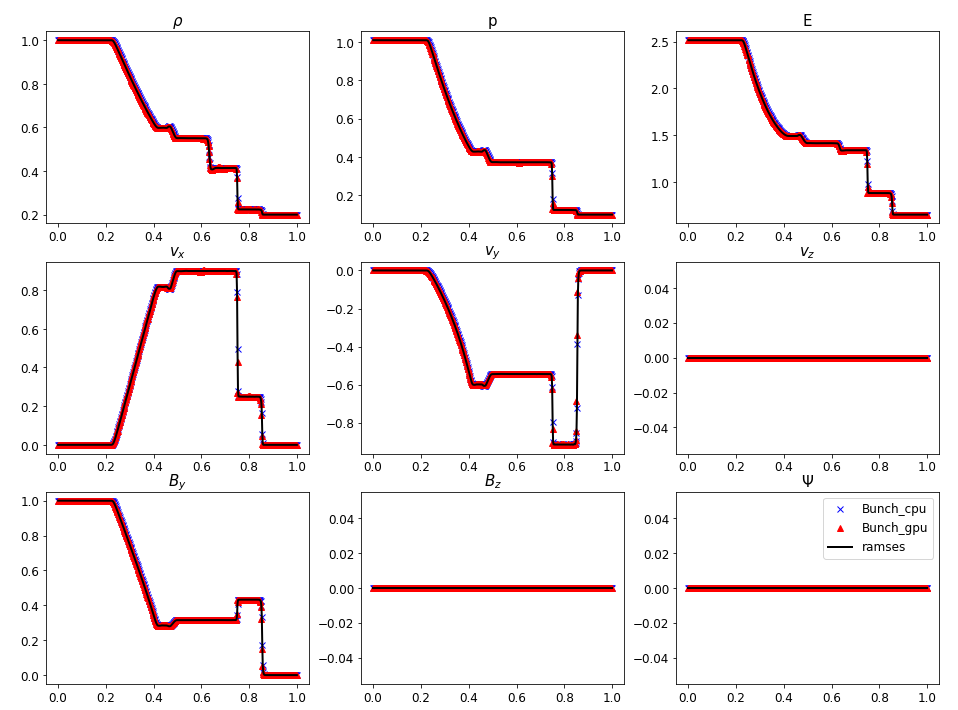}
\caption{Results of the Sod-Shock tube test for the Ramses, MHD\_Bunch CPU, and MHD\_Bunch GPU solvers. As the figure shows, the results are identical.}
\label{fig:1d_shock}       
\end{figure*}

The new \texttt{MHD\_Bunch} solver was validated using the Sod shock tube \citep{1d_shock} and the Orszag–Tang vortex experiment \citep{orszag-tang}. For the Sod shock tube test, both the \texttt{MHD\_Bunch} and \texttt{MHD\_Bunch\_CPU} solvers were tested against the existing RAMSES solver, \texttt{DISPATCH/RAMSES/HLLD}. The \texttt{MHD\_Bunch\_CPU} and \texttt{DISPATCH/RAMSES} tests used a single patch with 512 cells. The \texttt{MHD\_Bunch\_GPU} experiment used eight patches of 64 cells each, as the bunching scheduler is not developed for single-patch experiments. The results are shown in \FIGRef{fig:1d_shock}. As shown in the figure, all three solvers yield identical results, matching Fig. 4a in \citet{1d_shock}.

\begin{figure*}
    \centering
    \begin{subfigure}[b]{0.475\textwidth}
        \centering
        \includegraphics[width=\textwidth]{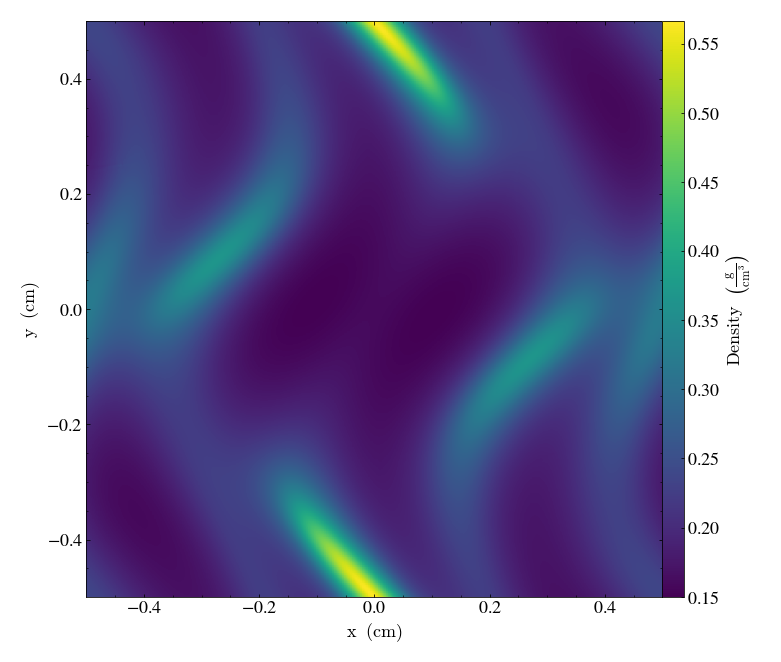}
        \caption{DISPATCH/RAMSES solution at time 0.1}
        \label{fig:OT_ram_0.1}
    \end{subfigure}
    \begin{subfigure}[b]{0.475\textwidth}
        \centering
        \includegraphics[width=\textwidth]{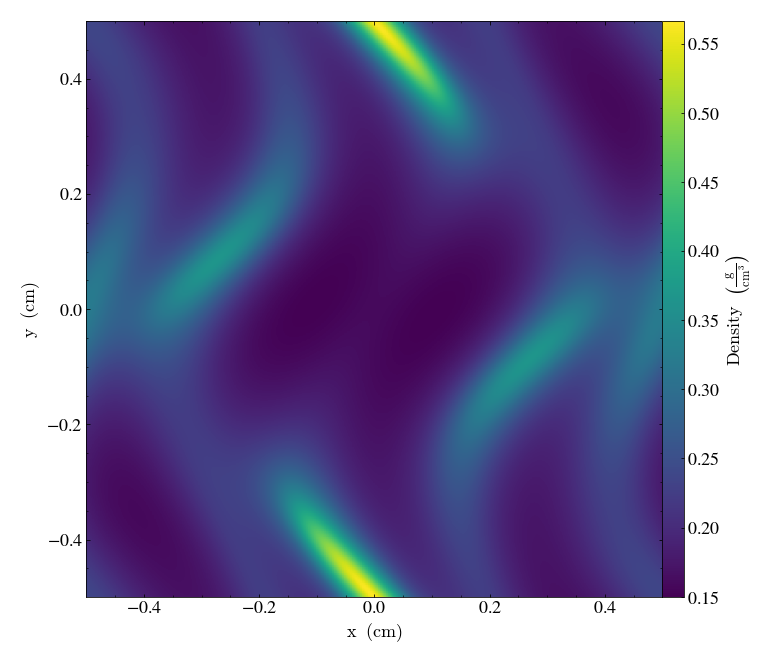}
        \caption{MHD\_Bunch solution at time 0.1}
        \label{fig:OT_gpu_0.1}
    \end{subfigure}
    
        \begin{subfigure}[b]{0.475\textwidth}
        \centering
        \includegraphics[width=\textwidth]{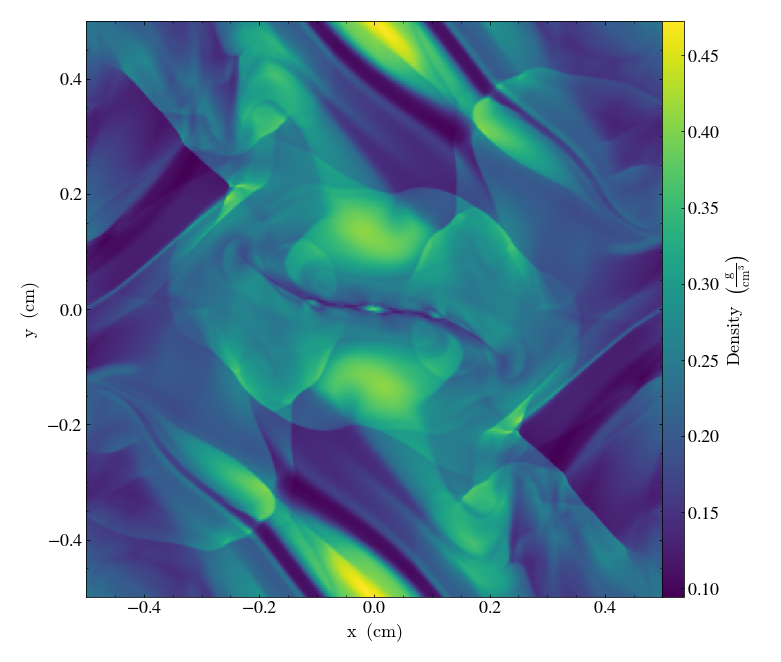}
        \caption{DISPATCH/RAMSES solution at time 0.6}
        \label{fig:OT_ram_0.5}
    \end{subfigure}
    \begin{subfigure}[b]{0.475\textwidth}
        \centering
        \includegraphics[width=\textwidth]{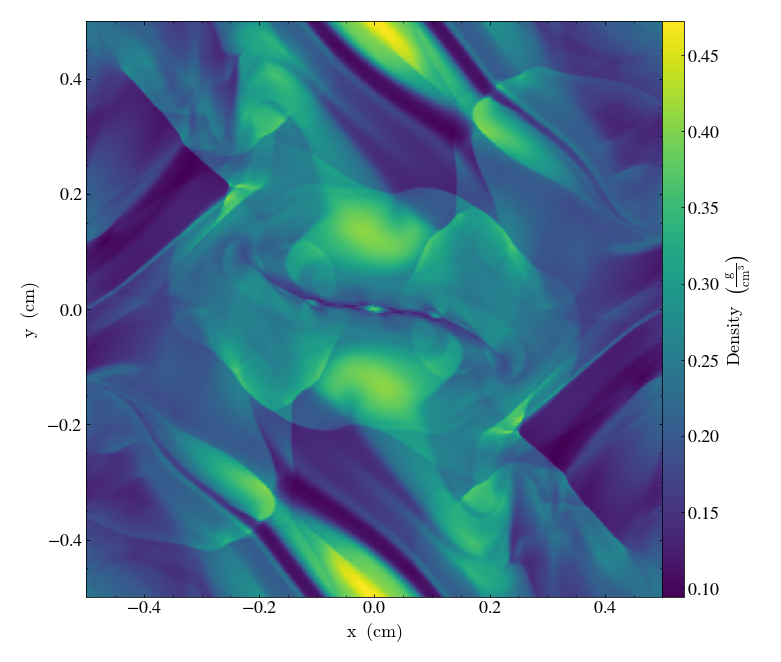}
        \caption{MHD\_Bunch solution at time 0.6}
        \label{fig:OT_gpu_0.5}
    \end{subfigure}

    \begin{subfigure}[b]{0.475\textwidth}
        \centering
        \includegraphics[width=\textwidth]{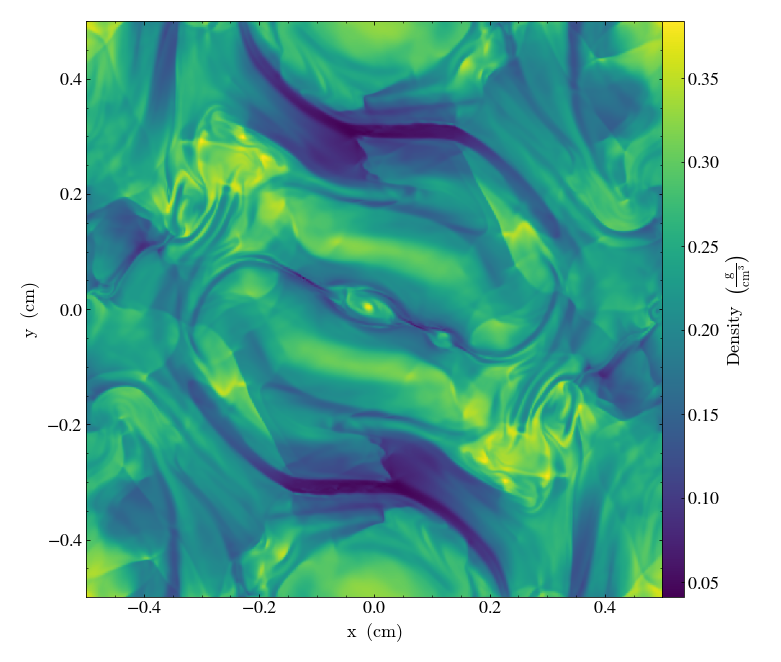}
        \caption{DISPATCH/RAMSES solution at time 1.0}
        \label{fig:OT_ram_1.0}
    \end{subfigure}
    \begin{subfigure}[b]{0.475\textwidth}
        \centering
        \includegraphics[width=\textwidth]{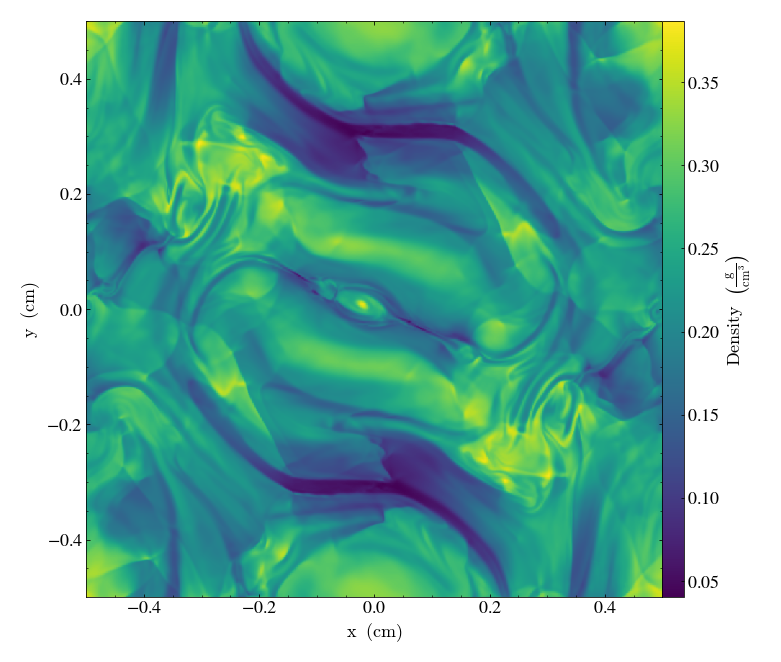}
        \caption{MHD\_Bunch solution at time 1.0}
        \label{fig:OT_gpu_1.0}
    \end{subfigure}
    \caption{Side by side snapshots of DISPATCH/RAMSES solution and MHD\_Bunch implementation running the Orszag-Tang experiment on a 512x512 grid. There is a slight mismatch in the later time steps.}
    \label{fig:OT_512}
\end{figure*}

The Orszag–Tang vortex experiment was run using a $512 \times 512$ grid. The results are shown in \FIGRef{fig:OT_512}. The output from the new \texttt{MHD\_Bunch} solver matches the existing \texttt{DISPATCH/RAMSES} solver up to approximately $t = 0.6$ s. Beyond this point, slight differences begin to appear. These variations do not originate from the solvers themselves, but from the exchange and interpolation of ghost zones. The Orszag–Tang experiment is a strictly symmetric setup, and any minor asymmetry can trigger a cascading divergence in the solution. Patches are updated in a semi-random order, meaning neighbouring cells may be updated at different times in different runs. This leads to slightly asynchronous patch times, resulting in differing ghost zone interpolations. Such minor discrepancies are sufficient to trigger deviations that grow visibly by $t = 1.0$ s. These differences arise not due to solver errors, but because of the inherent sensitivity of the symmetric initial conditions.

Validation of the \texttt{MHD\_Bunch} solver was performed for both GPU and CPU bunch execution. It is worth noting that similar small differences were also observed between independent runs using the \texttt{DISPATCH/RAMSES} solver, although they were more pronounced in the \texttt{MHD\_Bunch} case. With the solver validated, performance benchmarking was carried out on a larger test problem.

\subsection{\texttt{MHD\_Bunch} Performance}

To evaluate the performance of \texttt{MHD\_Bunch} in a realistic setting, we conducted a 3D turbulence test consisting of $16 \times 16 \times 16 = 4096$ patches, each with $24 \times 24 \times 24$ active cells, and $30 \times 30 \times 30$ cells including ghost zones. Simulations were executed on the astro04 system of the Tycho cluster\footnote{\url{https://wiki.nbi.ku.dk/tycho/Hardware}}, equipped with RTX A6000 GPUs. Once the simulation reached a steady state, performance was measured as the average number of patch updates per second during a 30-second interval.

We explored various configurations of GPU bunch sizes and CPU thread counts. A bunch size of 300 yielded the highest performance, and we found that at least 12 CPU threads per GPU were required to saturate the device and prevent GPU underutilisation between successive bunch updates.

\TABRef{tab:3d_results} presents the average execution time per cell and the corresponding relative speed-up. The GPU-accelerated implementation achieved a patch update rate approximately 110 times faster than the single-threaded CPU baseline (\texttt{DISPATCH/RAMSES}), inclusive of both kernel computation and host-device memory transfer. When normalised by the 12 CPU threads per GPU used to drive the GPU execution, this corresponds to the speed-up listed in the table. The CPU-based version of the new solver (\texttt{MHD\_Bunch\_CPU}) also demonstrated a notable improvement, achieving a 30\% speedup over the baseline.

The refined bunch scheduling algorithm ensures that the GPU remains active for more than 99.9\% of the wall time, significantly improving over the earlier mock-up implementation. In this benchmark, approximately one-third of the time is spent on data transfer between host and device, with the remaining two-thirds devoted to computation.

\begin{table*}
\centering
\begin{tabular}{||c c c c||} 
\hline
 & \texttt{DISPATCH/RAMSES} & \texttt{MHD\_Bunch\_CPU} & \texttt{MHD\_Bunch} \\ 
\hline\hline
Update time per cell (ns) & 179 & 137 & 18 \\
Speed-up                  & 1   & 1.3 & 9.8 \\
\hline
\end{tabular}
\caption{Average cell update time and relative speed-up for the 3D turbulence benchmark comprising 4096 patches with $24^3$ cells each. Measurements reflect steady-state performance averaged over 30 seconds.}
\label{tab:3d_results}
\end{table*}

\section{Discussion}
\label{sec:discussion}
The marked speedup observed in our preliminary mockup experiments necessitated a substantial refactoring of the original code, particularly the core \texttt{trace3D} function. This endeavour extended beyond incremental adjustments; it necessitated segmenting the function into smaller, more manageable components, refining parameter listings, reordering array indices, and re-sequencing operations to maximise execution efficiency.

This rigorous refactoring sheds light on the potential challenges of modernising other sophisticated codes with similarly complex core functions. It highlights the possibility that transitioning to GPU-centric programming environments, such as CUDA, may offer a more direct route to harnessing modern GPU computational power. On the other hand, if new codes are written with already small and manageable components--—a strategy that has other advantages in addition--—the effort required for OpenMP offloading can be considerably reduced. In our case, the process also made the CPU version more efficient, as evidenced by the performance gains in \TABRef{tab:3d_results}. With GPU offloading, we observed a 7.3-fold speedup in our initial mockup, comparing a single NVIDIA A100 GPU against seven CPU cores.

A key insight from our work is the necessity of adopting a hybrid view of scientific computing—one that goes beyond the binary dichotomy of CPU versus GPU. In DISPATCH's task-based method, computational roles are cleanly partitioned: the GPU handles physics update tasks, while the CPU performs pre-update and post-update tasks, including the interpolation between patches and data preparation. This division allows each architecture to play to its strengths. In the particular system we tested the code on, a 12-to-1 CPU-to-GPU ratio was sufficient to avoid GPU idle time, and remaining core capacity was available to either provide additional task updates, or else work on more complex pre- and post-processing, such as tabulated equation-of-state (EOS) lookups or adaptive physics modules-—without bottlenecking GPU execution. Such a hybrid strategy, balancing memory throughput and compute intensity, will be increasingly critical on emerging architectures.

The primary hurdle to achieving even greater performance lies in the memory transfer between the host system and the GPU. This bottleneck, due to limited support for asynchronous memory movement and pinned memory in the GCC compiler we used, constrains potential speedup. Asynchronous kernel execution is similarly unsupported, meaning only a single kernel can run on the GPU at any time. Some of these constraints are lifted in GCC 15, and more improvements can be expected over time. 
 In the work we report on here, we tested GCC versions from 10.2 to 14.1, facing the performance limitations mentioned. Alternative compilers, such as Clang and Intel, did not offer compatibility with the DISPATCH framework’s GPU offloading requirements and the available GPU hardware.

Our initial attempts to integrate the solver into the DISPATCH framework resulted in a notable performance drop. Specifically, the \texttt{MHD\_Bunch}'s update times slowed to ~92ns per cell from the ~20ns per cell observed in the HLLD mockup—making the full solver ~4.6 times slower than the mockup. However, by developing a strategy that uses three simultaneously active bunches--—one that pre-processes and bunches tasks, one that executes on the GPU, and one that post-processes results—--we could keep the GPU saturated, even including non-maskable transfer time, using only 8-12 CPU cores. This led to an overall speedup of 9.8, comparing one A6000 GPU to 12 CPU cores.

DISPATCH utilises asynchronous memory transfers to mask MPI-related latency efficiently. Extending this principle to host-GPU transfers could significantly reduce overhead, potentially improving performance by around 33\% in this case, and achieving a speedup close to a factor of 13.

The gains we observed, while not approaching the theoretical GPU-to-CPU FLOPS ratio, are consistent with performance enhancements reported by other teams \citep{Fusion_GPU,GenAsis}. For instance, GenASIS achieved a sixfold speedup---doubling with pinned memory---on Summit, and GENE achieved a 15-fold speedup using six GPUs for 42 CPU cores. These experiences highlight the significance of memory bandwidth as a performance bottleneck in memory-intensive applications, such as GenASIS, GENE, and DISPATCH. In this context, our performance aligns with the state of the art. A key advantage of DISPATCH, however, lies in its near-perfect MPI scaling. Unlike other frameworks where performance may degrade with increased parallelisation, DISPATCH maintains high efficiency across many ranks because communications are only needed between nearest-neighbour tasks and do not need to be globally synchronised. This ensures that the observed GPU speed-up can scale consistently across nodes, provided that each rank handles a sufficient workload to fully saturate the GPU. This robustness in strong scaling performance positions DISPATCH uniquely for exploiting modern heterogeneous HPC systems.

Recent advances in GPU memory capacities, which now approach those of CPUs, present new opportunities for optimisation. In future systems, the entire dataset may reside in GPU memory, reducing or eliminating the need for host-device transfer strategies. In such a scenario, MHD updates could be considered "free," enabling the integration of more complex physics without prohibitive computational costs. Our ongoing work includes the integration of a particle-in-cell (PIC) solver into DISPATCH to study particle acceleration in solar flares \citep{Haahr_2025}. The ability of each patch to dynamically switch between PIC and MHD modes—combined with efficient, GPU-accelerated MHD—will facilitate large-scale simulations previously deemed infeasible, offering a new frontier for multi-physics modelling in astrophysics.

\section{Conclusion}
\label{sec:conclusion}

Despite constraints imposed by OpenMP, our adaptation of the \texttt{MHD\_Bunch} solver for GPU utilisation delivered substantial performance improvements over CPU-only configurations. This success was enabled by a complete refactoring of some of the more extensive subroutines and loops in the original code, which improved both GPU compatibility and CPU efficiency. In benchmark tests, the mockup achieved a 7.3-fold speedup using a single GPU versus seven CPU cores. Within the integrated solver, patch update times improved by more than two orders of magnitude per CPU core, resulting in a total speedup of 9.8 when using 12 CPU threads. The CPU version also exhibited a 30\% performance improvement over its predecessor.

Beyond numerical gains, our results highlight the importance of hybrid GPU/CPU computing strategies in modern scientific codes. Rather than viewing the CPU and GPU as competing platforms, our approach partitions tasks according to architectural strengths: the GPU executes dense numerical updates, while the CPU simultaneously manages control flow, memory orchestration, and data preparation for the GPU, effectively hiding the cost of this work behind the GPU update cost. This hybrid model improves resource utilisation and provides flexibility to integrate complex or non-vectorisable physics modules into pre- and post-processing routines. 

Looking ahead, advances in compiler technology—--particularly support for asynchronous memory transfer and execution in GCC—--could unlock additional performance improvements. This, in combination with our hybrid model, will enable handling more complex simulations with greater efficiency, setting the stage for a new generation of large-scale astrophysical experiments.

In conclusion, our work demonstrates the viability and benefits of combining OpenMP-based GPU acceleration with modular CPU-GPU collaboration, establishing a practical, scalable and future-proof paradigm for high-performance computing in computational astrophysics.

\section*{Acknowledgements}
The Tycho HPC facility at the University of Copenhagen, supported by research grants from the Carlsberg, Novo, and Villum foundations, was used for carrying out the simulations, the analysis, and the long-term storage of the results.
TH acknowledges funding from the Independent Research Fund Denmark through grant No. DFF 4283-00305B.

During the preparation of this work we used ChatGPT in order to enhance the clarity of the writing. After using this tool, we reviewed and edited the content as needed and take full responsibility for the content of the publication.



\bibliographystyle{mnras}
\bibliography{refs} 








\bsp	
\label{lastpage}
\end{document}